\documentclass[prc,notitlepage, twocolumn,
showpacs,floatfix,nofootinbib,preprintnumbers,superscriptaddress,amsmath,amssymb,aps,longbibliography]{revtex4-2}

\usepackage[utf8]{inputenc}
\usepackage{graphicx}\usepackage{epsfig}
\usepackage{amsmath,amssymb}
\usepackage{bm}
\usepackage{color}
\usepackage{float}
\usepackage{dcolumn} 
\usepackage{enumerate}
\usepackage{multirow}
\usepackage{threeparttable}
\usepackage{here}
\usepackage{natbib}
\usepackage{ulem}
\usepackage[bookmarks=true,colorlinks,%
            citecolor=blue,linkcolor=blue,anchorcolor=blue,filecolor=blue,urlcolor=blue,%
           ]{hyperref}

\def\m@thcombine#1#2{%
  \setbox0=\hbox{$#1$}
  \setbox1=\hbox{$#2$}
  \ifdim\wd0>\wd1
    \setbox0=\hbox to\wd1{\hss\box0\hss}
  \else
    \setbox1=\hbox to\wd0{\hss\box1\hss}
  \fi
  \mathop{\vcenter{
    \offinterlineskip\box0\box1}}}
\def\lesim{\m@thcombine<\approx}
\def\gesim{\m@thcombine>\approx}

\begin{document}

\title{Exploring highly-deformed ground states involving the second intruder orbit in $Z>50$ even-even nuclei}
\author{Tsunenori Inakura}
\email{inakura@gmail.com}
\affiliation{Office of Institute Strategy, Institute of Science Tokyo, Tokyo 152-8550, Japan}

\author{Wataru Horiuchi}
\email{whoriuchi@omu.ac.jp}
\affiliation{Department of Physics, Osaka Metropolitan University, Osaka 558-8585, Japan}
\affiliation{Nambu Yoichiro Institute of Theoretical and Experimental Physics (NITEP), Osaka Metropolitan University, Osaka 558-8585, Japan}
\affiliation{RIKEN Nishina Center, Wako 351-0198, Japan}

\author{Shin'ichiro Michimasa}
\affiliation{RIKEN Nishina Center, Wako 351-0198, Japan}

\author{Masaomi Tanaka}
\affiliation{Faculty of Arts and Science, Kyushu University, Fukuoka 819-0395, Japan}

\date{\today}

\begin{abstract}

We present a systematic survey of even–even nuclei with $Z>50$ to identify where a very large prolate configuration driven by the second intruder orbit emerges. 
Within the energy density functional theory framework, we find in representative cases a pronounced prolate minimum at quadrupole deformation $\beta_2\approx$ 0.3--0.4. 
A characteristic feature of these minima is a local enhancement of the hexadecapole ($\beta_4$) component relative to nearby deformations, which is a clear fingerprint of the $\beta_2$--$\beta_4$ coupling expected for the second intruder orbit. 
Representative comparisons among three Skyrme interactions show a similar appearance of the highly deformed minimum and a local enhancement of $\beta_4$ at the prolate minimum, indicating qualitative robustness with respect to the interaction. The resulting maps highlight specific heavy nuclei where highly deformed ground states are anticipated.



\end{abstract}

\maketitle

\section{Introduction}

Nuclear shapes and their evolution across the chart provide key information on shell structure away from stability and govern bulk observables such as charge radii, moments, reaction cross sections, and fission barriers. 
Mapping where large deformations emerge—and which multipole components accompany them—provides benchmarks of energy density functional (EDF) descriptions.

In a recent analysis of the neutron-rich zirconium isotope $^{112}$Zr ($Z=40$, $N=72$)~\cite{Horiuchi23}, we demonstrated within a self-consistent EDF framework that the onset of partial occupancy of the second intruder orbit can stabilize a very large prolate shape. 
Under standard EDF conditions, the potential energy surface develops a pronounced minimum around quadrupole deformation $\beta_2 \approx 0.4$.
These results motivate a systematic assessment of how generic this intruder-driven mechanism is, and how robust it remains with respect to the effective interaction in heavier even–even systems.

Quadrupole–hexadecapole ($\beta_2$-$\beta_4$) shapes affect bulk observables such as charge radii and reaction cross sections, and can also influence the outer part of fission paths.
It is well established that the $\beta_{2}$–$\beta_{4}$ coupling is strong; pure $\beta_{4}$ minima are exceptional, while sizable $\beta_{4}$ components typically accompany large quadrupole deformation~\cite{Delaroche10,Kumar23}.
Recent beyond-mean-field studies report that these $(\beta_2,\beta_4)$ trends are robust under reasonable interaction changes, and surveys in heavy systems likewise stress the role of higher multipoles along fission trajectories~\cite{Kumar23,Guzman20,Okada23,Guzman25}.

Guided by the $^{112}$Zr case, we survey even–even nuclei with $Z>50$ with the aim of exploring where the second intruder mechanism actually materializes. 
In representative nuclei within this region the calculated deformation landscapes exhibit a pronounced prolate minimum at $\beta_{2}\!\approx$ 0.3$–$0.4.
A characteristic feature at the minimum is an enhanced hexadecapole component relative to its surroundings, providing a clear manifestation of the $\beta_{2}$–$\beta_{4}$ coupling. 
The present maps indicate where highly-deformed ground states are expected in heavy systems and what $\beta_{4}$ response they carry, while keeping the scope intentionally simple (axial symmetry, a common HO basis truncation, no triaxial or octupole degrees of freedom).

The paper is organized as follows. Section~\ref{Method}  summarizes the common setup used in all calculations.
Section~\ref{Results} presents the systematics and the interaction dependence, including an overlay figure that illustrates the robustness of the large prolate minimum across SkM$^*$, SLy4, and UNEDF1. 
Section~\ref{Summary} concludes with the main findings, scope, and limitations.

\section{Method}
\label{Method}

We perform a systematic calculation of ground-state properties
of even-even nuclei from Te $(Z=52)$ to Hg $(Z=80)$
isotopes by the Skyrme Hartree-Fock-Bogoliubov (HFB) model
using the HFB solver, HFBTHO~\cite{Stoitsov13}.
Deformation energy curves are obtained with an axial HFB calculation. The quadrupole moment $Q_{20}$ (i.e., $\beta_2$) is constrained by a standard quadratic term, and at each fixed 
$\beta_2$ the HFB energy is minimized. Other multipoles are not constrained in this work.
The single-particle wave functions are expanded in an axially deformed harmonic-oscillator basis including all major shells up to $N_\mathrm{max}=14$. 
We find this size adequate to describe the global systematics discussed below. 
For some selected heavy nuclei we also increased the basis and saw only modest changes in bulk observables. The same truncation is used for all nuclei.
At each constrained value of $\beta_2$, we adjust the aspect ratio of the axially deformed HO basis to approximately follow the target deformation, 
in order to reduce basis–shape mismatch. 
No nucleus-specific tuning is applied.
The SkM$^\ast$ interaction is employed~\cite{Bartel82}
as it is known to reasonably describe nuclear deformations.
The SLy4~\cite{Chabanat98} and UNEDF1~\cite{Kortelainen12} interactions have been used to check the dependence of the results on the interaction considered in the calculations.
A standard mixed-type pairing interaction~\cite{Sandulescu05} is used. 
In this work, we use only axial HFB solutions. 
We focus on global trends such as the rms radii, diffuseness, reaction cross sections, 
which are dominated by the bulk deformation and density profile. 
In many even–even nuclei, the change in the rms radius when triaxiality is allowed is usually 
modest compared with spectroscopic effects. 
Large-scale EDF studies that include triaxial shapes reproduce charge-radius systematics well and support this view~\cite{Delaroche10,Nomura21}. 
At the same time, there are known $\gamma$-soft regions (e.g., Os–Pt) where triaxiality can be relevant, and recent studies in Ru isotopes also indicate that triaxiality may affect radius~\cite{Maass25}.
Therefore, our axial results should be seen as a simple trend map.

\section{Numerical Results}
\label{Results}

\subsection{Exploring the characteristic deformation in $Z>50$ even-even nuclei}

\subsubsection{Heavy mass region}  

\begin{figure}[tb]
\begin{center}
\includegraphics[width=\columnwidth,keepaspectratio]{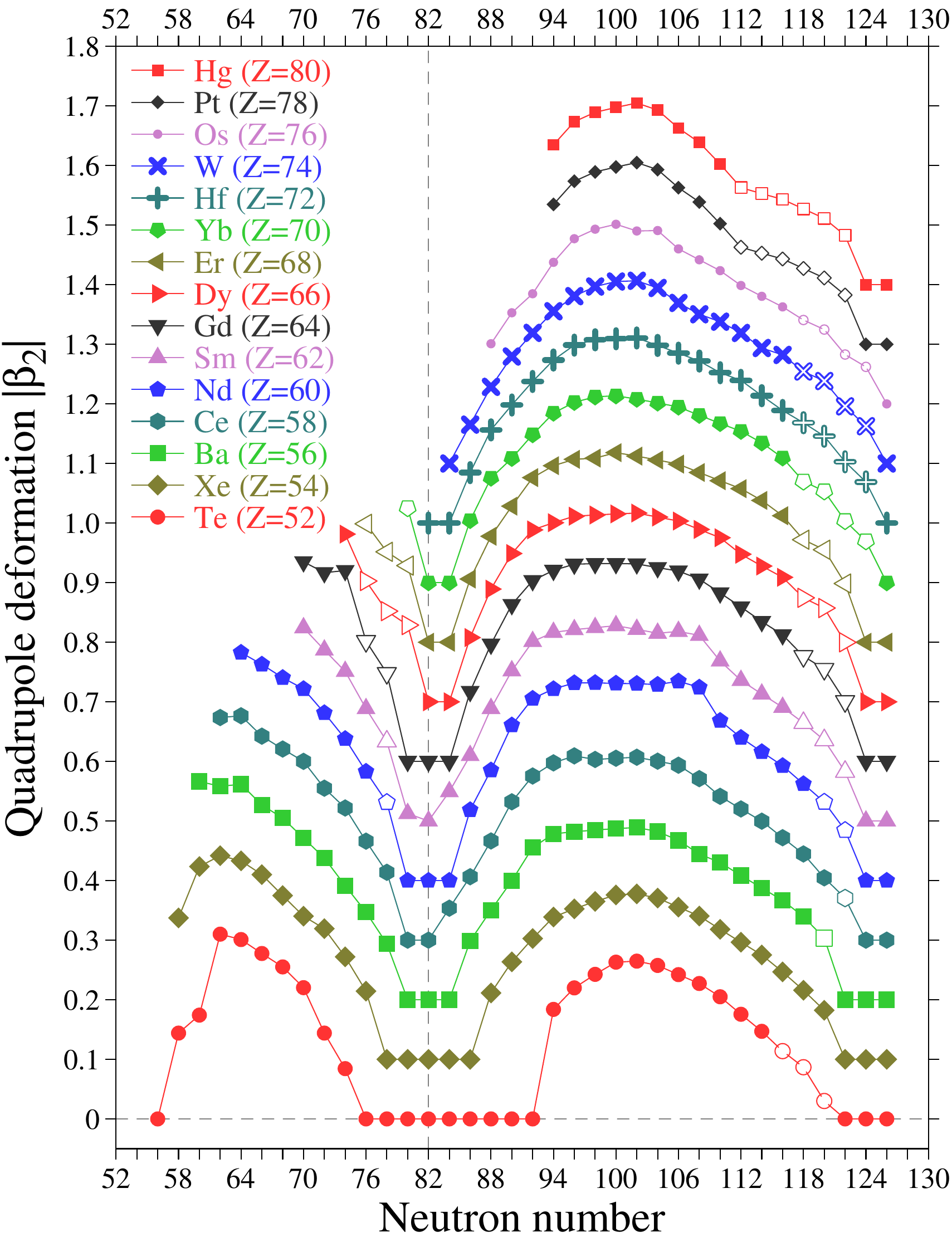}
\caption{Quadrupole deformation parameters $|\beta_2|$ of even-even nuclei from proton dripline to $N=126$ of Te ($Z=52$) to Hg ($Z=80$) isotopes as a function of the neutron number. 
For visibility, the $|\beta_2|$ values of each isotope is shifted by $0.05\times (Z-52)$ and oblate shapes ($\beta_2<0$) are represented by open symbols. 
}
\label{isotopes.all}
\end{center}
\end{figure}

Figure~\ref{isotopes.all} displays the quadrupole deformation parameters
$|\beta_2|$ of even-even nuclei from Te ($Z=52$) to Hg ($Z=80$) isotopes
as a function of the neutron number from proton dripline to $N=126$.
Open symbols denote oblate deformation ($\beta_2<0$). 
All isotopes are spherical at $N=82$
and show gradual enhancement of the $\beta_2$ values for $N>82$.
The deformation grows with increasing $N$ by the neutron occupation
of the intruder [550]1/2 orbit coming from the spherical $0h_{11/2}$
orbit, and the $\beta_2$ values reach up to $\approx 0.3$ at $N \approx 100$.
For $N\gtrsim 100$, the deformations of the ground states become smaller,
and change to oblate shapes. 
At $N \approx 126$, the ground states become spherical again.
While all isotopes have similar behaviors of the deformation
of the ground states, only Nd ($Z=60$) and Sm ($Z=62$) isotopes
exhibit kinks of the $\beta_2$ values at $N=106$ and 108.
The proton number 60 may have some relations
with the onset of the large deformation in $^{100}$Zr with the neutron number 60 \cite{Ansari17}.

\begin{figure}[tb]
\begin{center}
\includegraphics[width=\columnwidth,keepaspectratio]{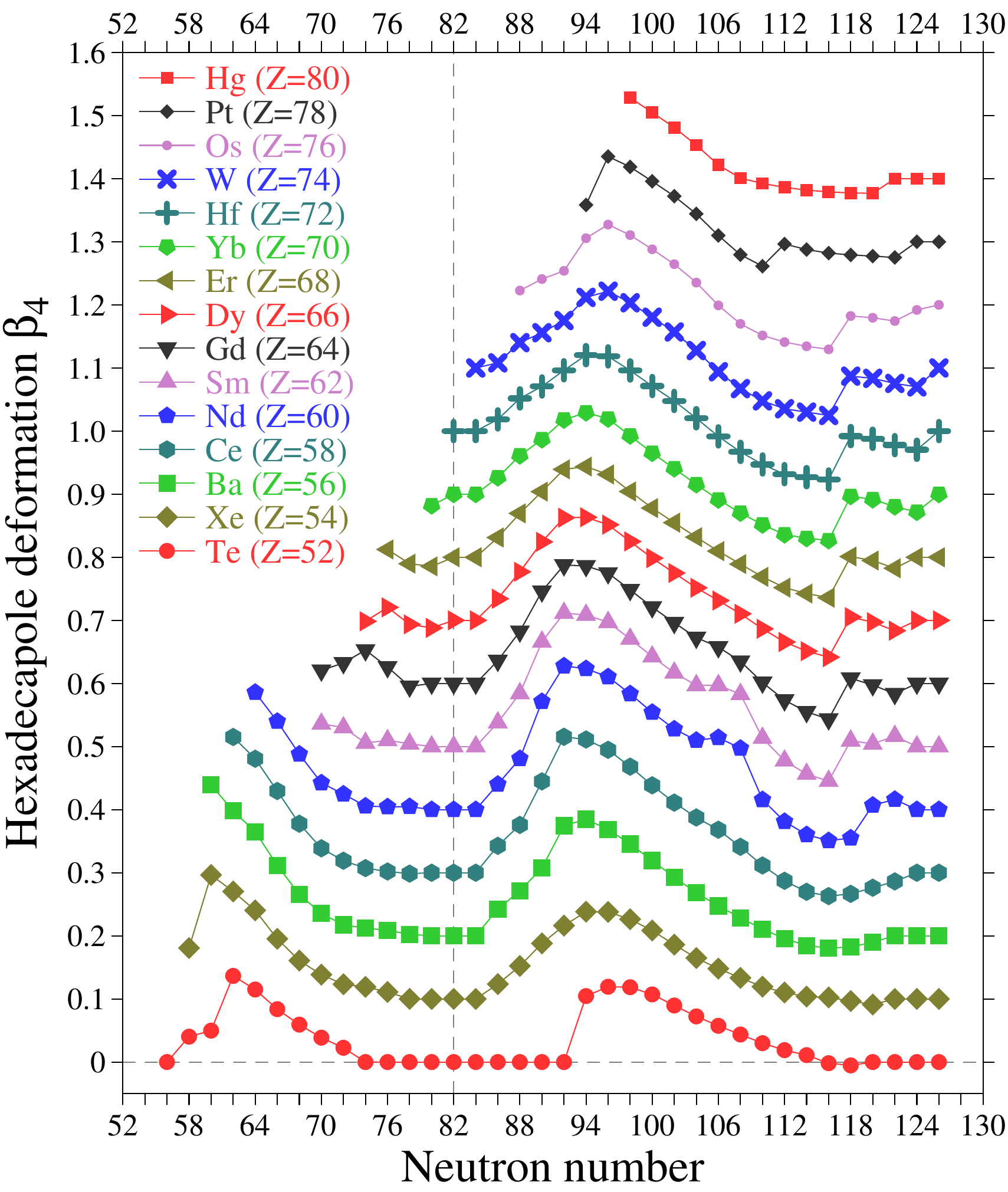}
\caption{Hexadecapole deformation parameters $\beta_4$ of even-even nuclei from proton dripline to $N=126$ of Te ($Z=52$) to Hg ($Z=80$) isotopes as a function of the neutron number. 
For visibility, the $\beta_4$ values of each isotope is shifted by $0.05\times (Z-52)$. 
}
\label{beta4.all}
\end{center}
\end{figure}

The hexadecapole deformation $\beta_4$ serves an indicator of
the occupation of the second intruder orbit as it is enhanced by the occupation of
an elongated orbit \cite{Horiuchi23,Horiuchi22}.
Figure~\ref{beta4.all} displays the calculated $\beta_4$ values
of the isotopes.
The $\beta_4$ values, linking to $\beta_2$, increase gradually for $N>82$ due to the fractional neutron occupation of
the intruder [550]1/2 orbit, 
reaching maximum values at $N\approx 90$.
Following this peak, the $\beta_4$ values decrease gradually and turn negative values, approaching to zero at $N\approx 126$.
In several isotopes, the hexadecapole deformations jump at $N \approx 120$, 
which corresponds to the shape changes from prolate to oblate deformations shown in Fig.~\ref{isotopes.all}. 
Similar to $\beta_2$, there are bumps of $\beta_4$ in Nd and Sm isotopes at $N=106$ and 108.
Therefore, $^{166,168}$Nd and $^{168,170}$Sm likely have the ground states involving the second intruder orbit.

\begin{figure}[htb]
\begin{center}
\includegraphics[width=\columnwidth,keepaspectratio]{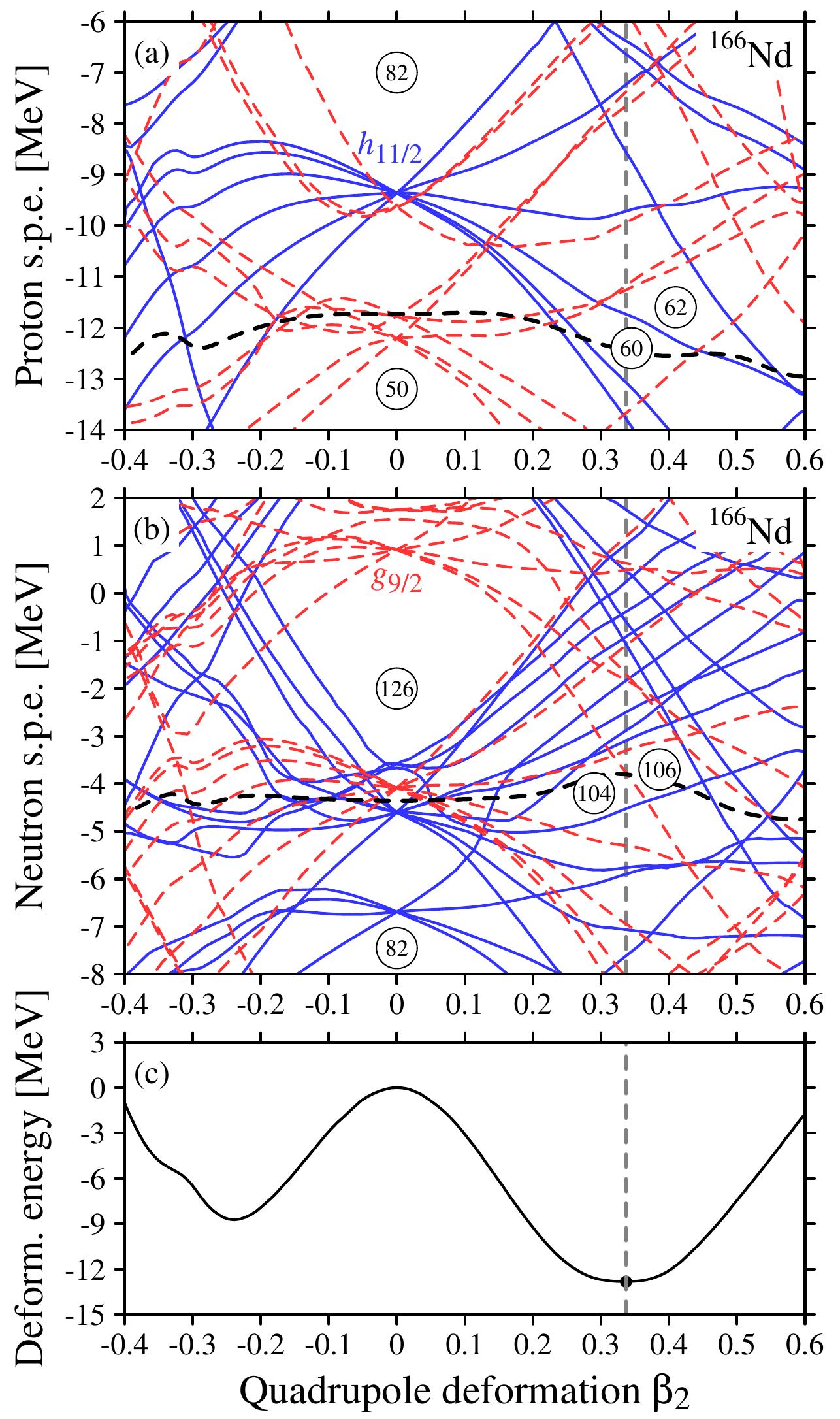}
\caption{(a) Proton and (b) neutron single-particle energies (s.p.e) and (c) deformation energies of $^{166}$Nd
  as a function of quadrupole deformation parameter $\beta_2$. 
  Blue solid and red dashed curves in the panels (a) and (b)
present negative and positive parity single-particle levels, respectively. 
Thick dashed curves denote the proton and neutron chemical potentials. 
A vertical dotted line indicates the $\beta_2$ value of the ground state for a guide to the eye.
}
\label{pec.config.166Nd}
\end{center}
\end{figure}

We plot in Fig.~\ref{pec.config.166Nd} the proton and neutron single-particle energies of $^{166}$Nd as a function of the quadrupole deformation parameter $\beta_2$. 
  The deformation energy, which is measured from that with the spherical shape ($\beta_2 = 0$), is also plotted as a guide.
  We see that the deformation energy of $^{166}$Nd 
 has the minimum at $\beta_2 = 0.33$, showing large quadrupole deformation.
  In $^{166}$Nd, the asymptotic quantum number of the most elongated orbit is [660]1/2,
  which comes from the spherical $0i_{13/2}$ orbit.
Since the nuclear deformation is so large, the elongated, next intruder orbit with $[651]1/2$ coming from the spherical $1g_{9/2}$ orbit is occupied
with an occupation probability 0.48.
The hexadecapole deformation parameter is also large, $\beta_4=0.11$.
This second intruder orbit plays a role in forming
the shell gaps at $N = 106$, and hence such a large deformation is stabilized. 
Note that proton single-particle energies 
also has a large shell gap with $Z=60$ at large deformation, 
occupying the enlarged orbits with $[550]1/2$ and $[541]3/2$. 
Both neutrons and protons favor the large deformation.

Recent beyond-mean-field studies show that $\beta_2$ and $\beta_4$ are strongly linked.
In the rare-earth region (Sm and Gd), a Gogny HFB+GCM calculation finds that including $\beta_4$ dynamics gives extra correlation energy of a few hundred keV, and produces low-lying $0^+$ states with mixed $(\beta_2,\beta_4)$ character~\cite{Kumar23}.
For heavy nuclei (Ra--Pu), a microscopic study reports a similar coupling pattern, a small but stable negative $\beta_4$ region near $N\!\approx\!184$ even after zero-point motion, and a change of the coupling strength along the isotopic chains~\cite{Guzman25}.
These works also show that the minimum can shift when $\beta_4$ is included.
Our mean-field maps are consistent with these general trends.
Here we provide a broad systematics and the link to second intruder occupation while a full GCM analysis is outside the scope of the present survey.

At the prolate $\beta_2$ minimum, the “depth’’ along $\beta_4$ can be quantified by the curvature of the collective potential with respect to 
$\beta_4$, i.e. the second derivative $\partial^2 E/\partial\beta^2_4$.
The coupling between $\beta_2$ and $\beta_4$ is captured by the cross second derivative $\partial^2 E/\partial\beta_2\partial\beta_4$.
Earlier beyond–mean–field analyses~\cite{Kumar23,Guzman25} show that such $\beta_2-\beta_4$ coupling can be sizable and 
may shift minima by up to the MeV scale in selected isotopic chains. In this paper, $\beta_4$ is used mainly as a global indicator and 
we focus on systematic trends. A dedicated two–dimensional $(\beta_2,\beta_4)$ study for a few benchmark nuclei, 
including explicit evaluations of these curvatures, will be pursued next.

\subsubsection{Proton-rich region}

\begin{figure}[!tb]
\begin{center}
\includegraphics[width=\columnwidth,keepaspectratio]{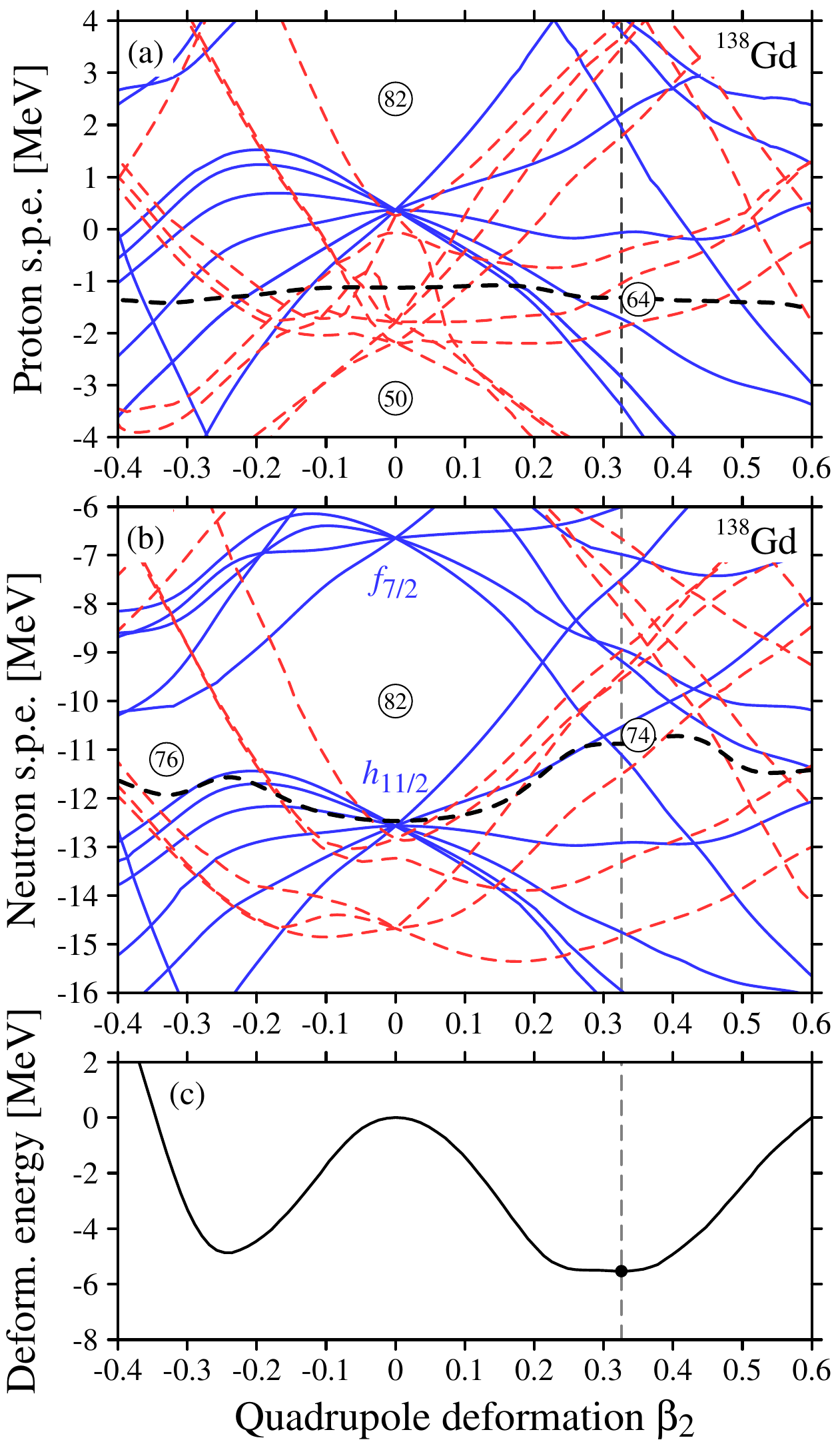}
\caption{Same as Fig.~\ref{pec.config.166Nd} but for $^{138}$Gd.
}
\label{pec.config.138Gd}
\end{center}
\end{figure}

 We also find a contribution of the second intruder orbit
       in a proton-rich nucleus $^{138}$Gd with $Z=64$ and $N=74$
       as indicated by the kink behavior seen in Figs.~\ref{isotopes.all} and ~\ref{beta4.all}.
  The deformation energy, proton and neutron single-particle levels
  of $^{138}$Gd are shown in Fig.~\ref{pec.config.138Gd}.
  Calculated $^{138}$Gd is highly deformed, with the energy minimum at $\beta_2=0.33$ and $\beta_4=0.05$.
  In the ground state of $^{138}$Gd, the second intruder neutron orbit originating 
  from the spherical $1f_{7/2}$ orbit is occupied with an occupation probability 0.73,
  following the same mechanism in neutron-rich nuclei $^{112,114}$Zr~\cite{Horiuchi23}.
  Additionally, the proton shell structure shows a gap with $Z=64$ at large $\beta_2$.
We remark that the ground state of $^{138}$Gd is observed,
  showing a large quadrupole deformation~\cite{Procter11}.

\begin{figure}[htb]
\begin{center}
\includegraphics[width=\columnwidth,keepaspectratio]{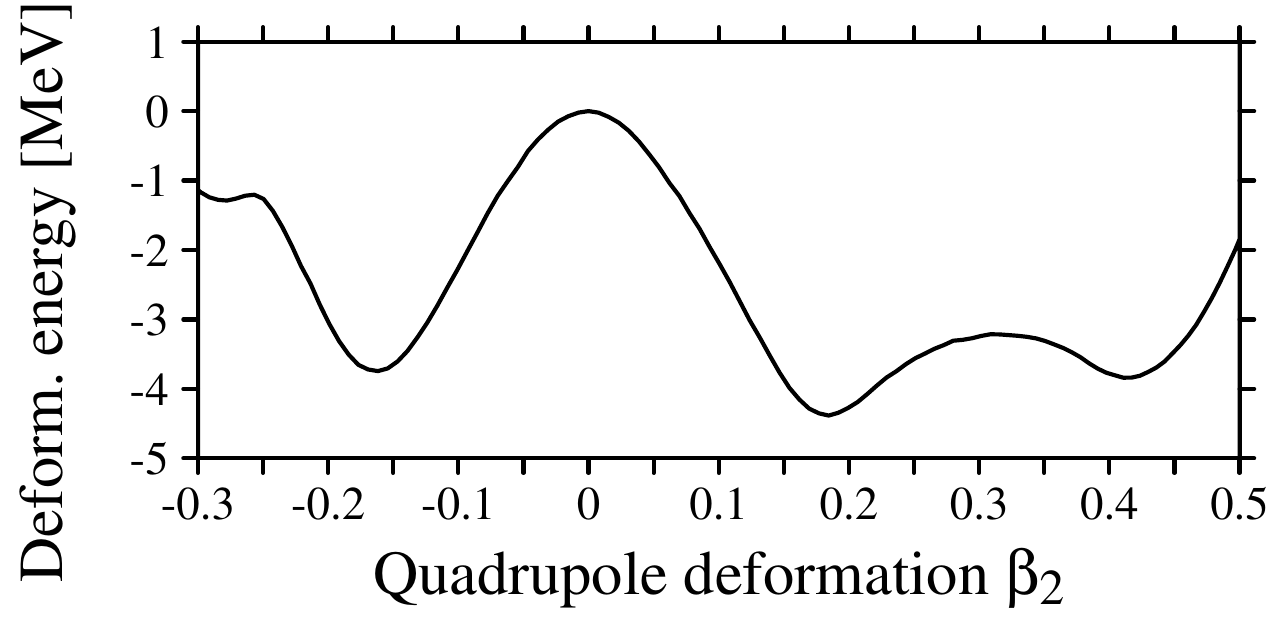}
\caption{Deformation energy curve for $^{272}$No obtained from axial HFB calculations with the HFBTHO solver, using an axially deformed harmonic-oscillator basis truncated at $N_{\max}=14$. Octupole deformation ($\beta_3$) is not included.
}
\label{pec.272No}
\end{center}
\end{figure}

\subsubsection{Case of $^{272}$No}

The deformation energy curve for $^{272}$No is obtained from axial HFB calculations using an axially deformed HO basis truncated at $N_{\max}=14$. 
For such a heavy neutron-rich nucleus, the depths of the minima can change when we use a larger basis. 
We tried a larger basis and saw only modest changes in bulk values (energy differences along the curve, $\beta_2$, $\beta_4$, rms radius), so the global picture stays the same. 
We still find a prolate minimum near $\beta_2 \approx 0.4$ connected to the second intruder. 
This minimum looks robust, but we cannot decide if it is the ground state or an excited one within the present model; adding $\beta_3$ and a larger basis could change the ordering. 
Octupole shapes ($\beta_3$) are important in actinides and superheavy nuclei and often lower the third barrier and modify a shallow third minimum~\cite{Guzman20,Okada23}.

\subsection{Observables}

\begin{figure}[!tb]
\begin{center}
\includegraphics[width=\columnwidth,keepaspectratio]{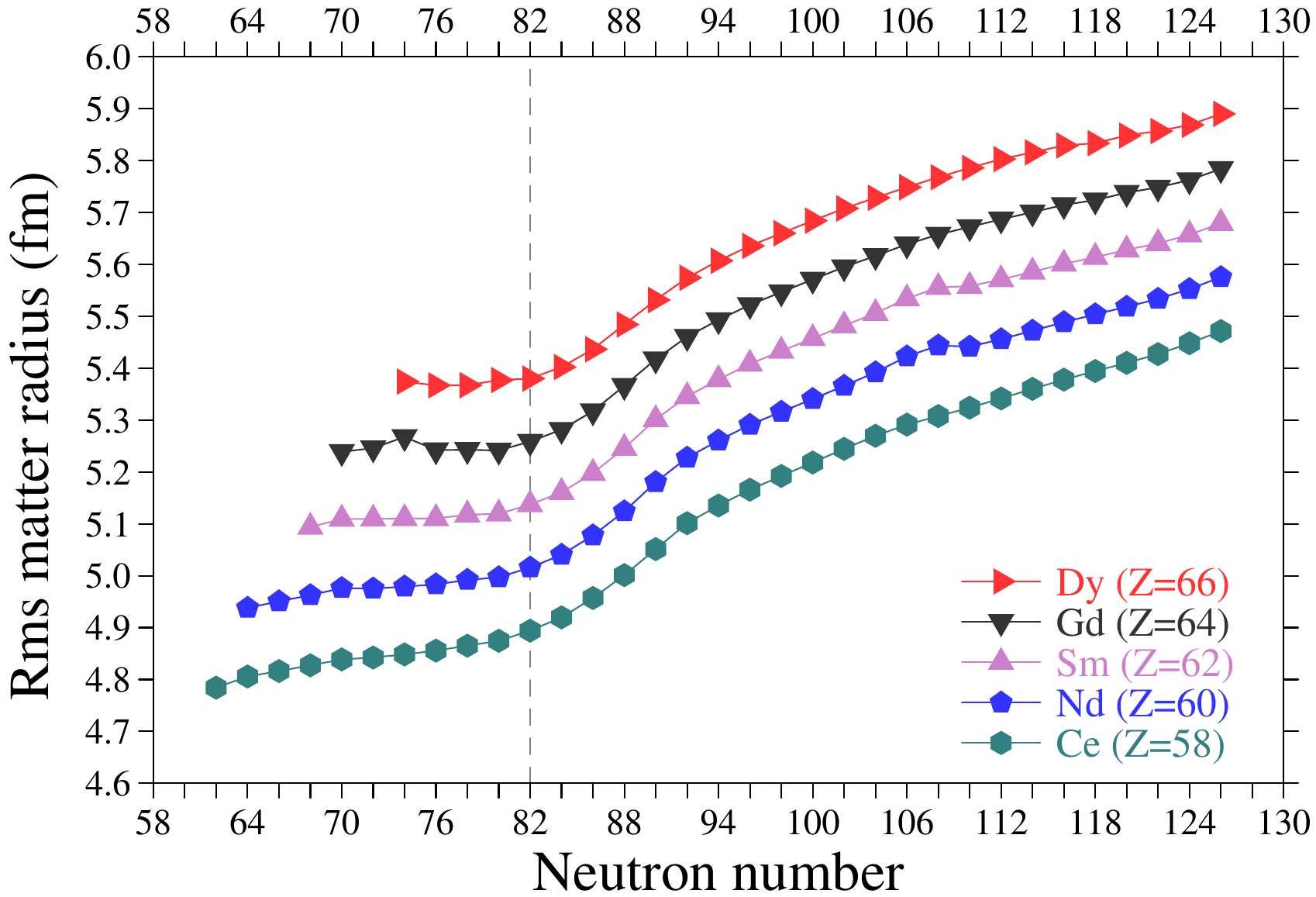}
\caption{Matter radii of Ce, Nd, Sm, Gd, and Dy isotopes
  as a function of neutron number. For visibility, the values are shifted by
$0.05\times(Z-58)$ fm.}
\label{radii.fig}
\end{center}
\end{figure}

Here we discuss possible observables to identify this characteristic
deformation.

The occupation of the second intruder orbit
is reflected in the density profiles near the nuclear surface.
Figure~\ref{radii.fig} plots the matter radii of these isotopes.
We see kink behavior at $N\approx 106$ for Nd and Sm isotopes
although the changes seem to be small.

\begin{figure}[!tb]
\begin{center}
\includegraphics[width=\columnwidth,keepaspectratio]{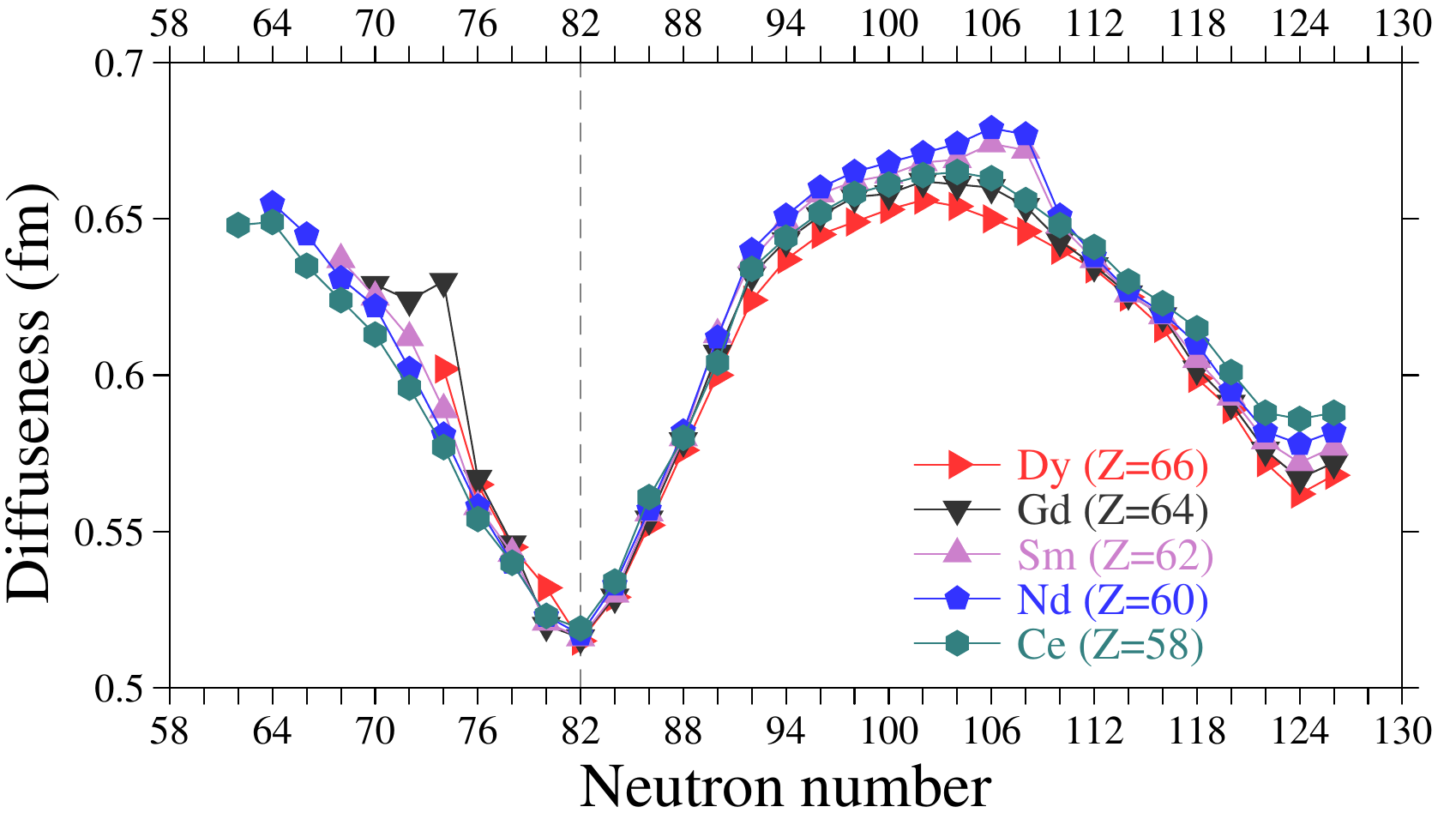}
\caption{Diffuseness parameters of Ce, Nd, Sm, Gd, and Dy isotopes
  extracted from these matter density distributions
  as a function of neutron number.}
\label{diff.fig}
\end{center}
\end{figure}

This difference becomes more apparent by analyzing
the diffuseness of the matter density distribution $\rho(r)$
at the surface region~\cite{Hatakeyama18}.
Assuming the model density distribution
of the two-parameter Fermi (2pF) function
$\rho_{\rm 2pF}(r)=\rho_0\{1+\exp[(r-R)/a]\}$.
The radius $R$ and diffuseness parameters $a$ are obtained by
minimizing~\cite{Hatakeyama18}: $\int_0^{\infty}dr\, r^2\left|\rho(r)-\rho_{\rm 2pF}(r)\right|$.
Figure~\ref{diff.fig} draws the diffuseness parameter extracted
from the matter density distribution obtained
by the present HFB calculations.
Apparently, the nuclear surface becomes more diffused
and produces the kink behavior in the $a$ values
when the elongated second intruder orbit is occupied at $N=70$ and 106.
As shown in Ref.~\cite{Hatakeyama18},
the diffuseness can be determined
by measuring the medium- to high-energy
proton-elastic scattering cross sections at the first peak position.
It is desirable to systematically measure these cross sections
to detect the structure change, which is very unique in
a isotopic chain of the highly deformed states.

Another possibility is to measure the total reaction or interaction cross sections at medium to high incident energies.
The total reaction and elastic scattering
cross sections are useful to study
the size properties of unstable nuclei~\cite{Tanihata85}.
The cross section calculations are well established
and can be obtained by using a high-energy reaction theory,
the Glauber theory~\cite{Glauber}.
We use the same setup to calculate the total reaction cross section
as Refs.~\cite{Horiuchi22, Horiuchi23}.
The inputs to the theory is the nuclear density distributions
and the profile function which reproduces
the nucleon-nucleon scattering properties.
A standard parameter set of the profile function
is used and which is tabulated in Ref.~\cite{Ibrahim08}.
For a carbon target, we employ
a harmonic-oscillator type density distribution reproducing
the experimental charge radius.
It should be noted that no {\it ad hoc} parameter
is introduced after all the inputs are set.
The validity of this approach has been confirmed
in many examples of medium- to high-energy nuclear reactions
involving unstable nuclei~\cite{Horiuchi10, Horiuchi12, Horiuchi15, Horiuchi16, Nagahisa18}.

\begin{figure}[!tb]
\begin{center}
\includegraphics[width=\columnwidth,keepaspectratio]{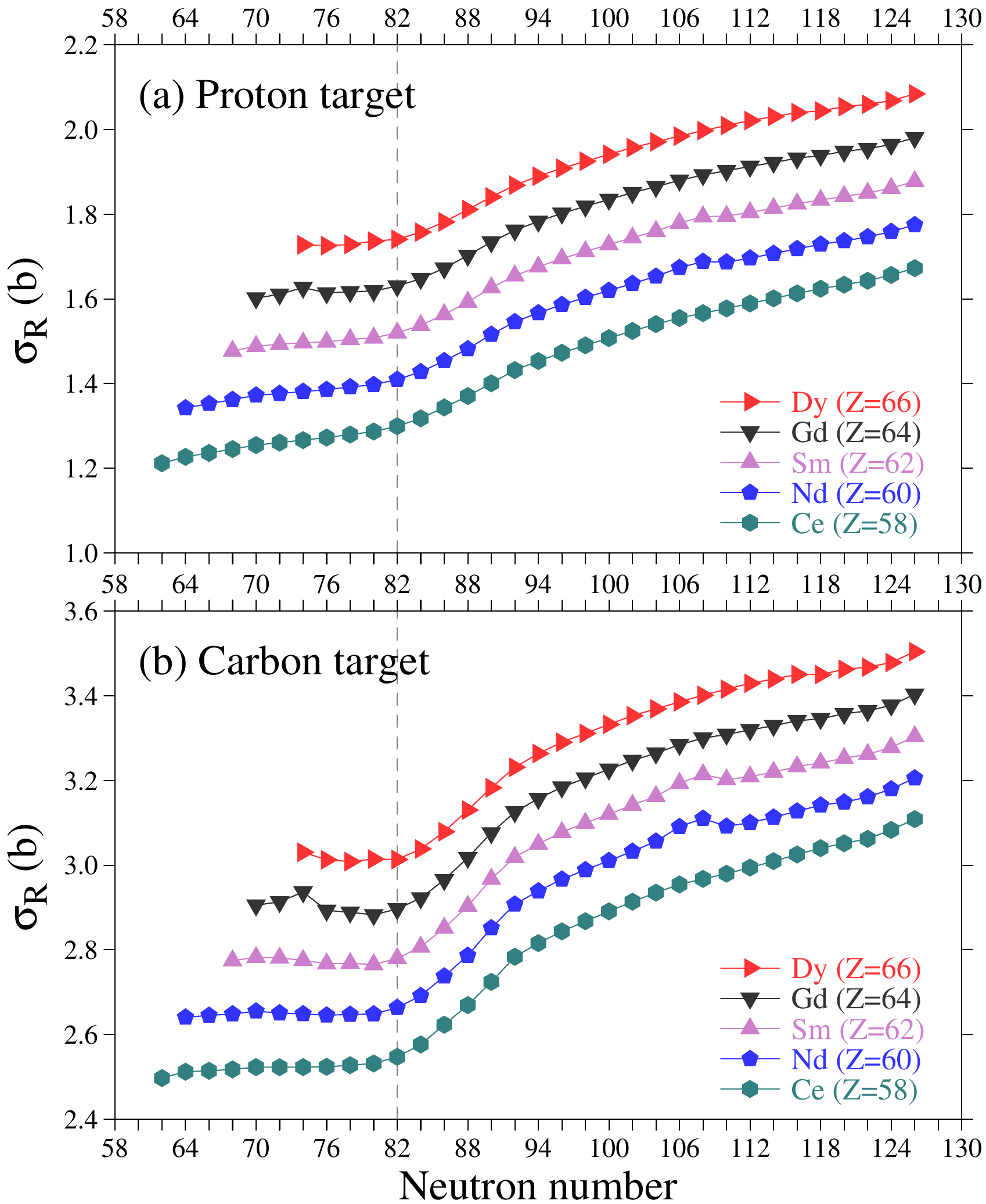}
\caption{Total reaction cross sections
  of Ce, Nd, Sm, Gd, and Dy isotopes on (a) proton and (b) carbon
  targets at 300 MeV/nucleon as a function of neutron number.
  For visibility, the cross sections are shifted by $0.05\times(Z-58)$.
}
\label{rcs.fig}
\end{center}
\end{figure}

It is known that the size properties of
the nuclear ground state is well reflected in the total reaction
cross section at a few to several hundred MeV/nucleon.
To discuss the feasibility of detecting the kink due
to the occupation of the second intruder orbit,
we compute the total reaction cross sections using the density
distributions obtained by the HFB model.
Note that the density distribution in the HFBTHO code
is the intrinsic one and
is converted to the laboratory frame density $\rho(r)$ by 
averaging the intrinsic density distribution
over angles as prescribed in Ref.~\cite{Horiuchi12}.
Figure~\ref{rcs.fig} plots the calculated total reaction cross sections on
proton and carbon targets. Incident energy is chosen
at 300 MeV/nucleon as it is a typical incident energy at RIKEN~\cite{Yano07}.
For a proton target, the cross section behavior is similar to
that of the matter radii, while for a carbon target,
the kink behaviors around $N=70$ and 106 are more pronounced.
The above observations can be understood by reminding that
the cross section on a proton target well reflects the change
of the nuclear radii but is less sensitive
to the surface part of the nuclear density distribution
than that on a carbon target~\cite{Horiuchi14}.
In sense of the sensitivity to the nuclear surface,
a nuclear target is in general more advantageous than the proton target
to probe the kink behavior due to the occupation of the second
intruder orbit.

\subsection{Interaction dependence}

\begin{figure}[!tb]
\begin{center}
\includegraphics[width=0.490\textwidth,keepaspectratio]{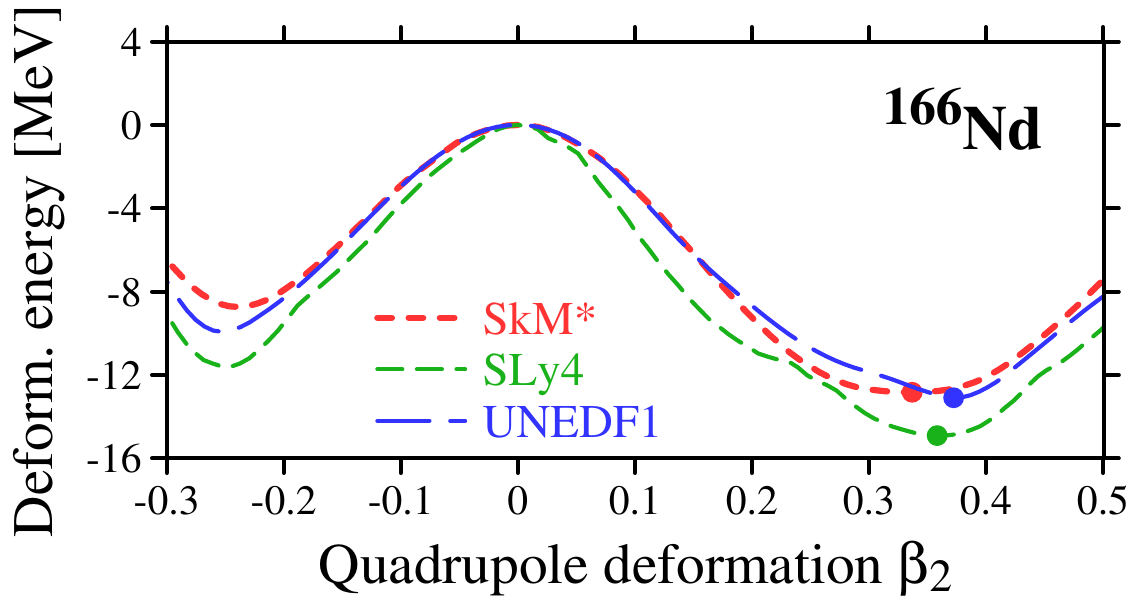}
\caption{Axial-HFB (HFBTHO) deformation-energy curve of $^{166}$Nd.
Energies are shifted to the spherical point ($\beta_2$=0). SkM*, SLy4, and UNEDF1 give the same large prolate minimum at $\beta_2 \approx$ 0.35; differences are small. A curvature change around $\beta_2 \approx$ 0.3 indicates the emergence of the second-intruder minimum.}
\label{interactionDep}
\end{center}
\end{figure}

We repeated the calculations with three Skyrme interactions (SkM*, SLy4, UNEDF1) using the same setup.
Figure~\ref{interactionDep} shows a representative case ($^{166}$Nd) in the second-intruder region.
SkM*, SLy4, and UNEDF1 produce a common large prolate minimum at $\beta_2\approx$ 0.35.
The curves almost overlap, indicating only modest differences in depth and local stiffness.
We also note a small change of curvature around $\beta_2 \approx$ 0.30, consistent with the onset of the second-intruder configuration.
Together, these features demonstrate that the appearance of the highly-deformed ground state is robust against the choice of Skyrme interaction.
At the prolate minimum, the extracted $\beta_4$ is similar among the three interactions, and the three curves overlap closely in a small window around the minimum, suggesting a qualitatively robust $\beta_2$–$\beta_4$ coupling (see Sec.~III~A).
This interaction-robust behavior agrees with a Gogny HFB+GCM analysis in the rare-earth region, which found stable $(\beta_2,\beta_4)$ trends under reasonable interaction changes~\cite{Kumar23}.

\begin{figure}[htb]
\begin{center}
  \includegraphics[width=\columnwidth,keepaspectratio]{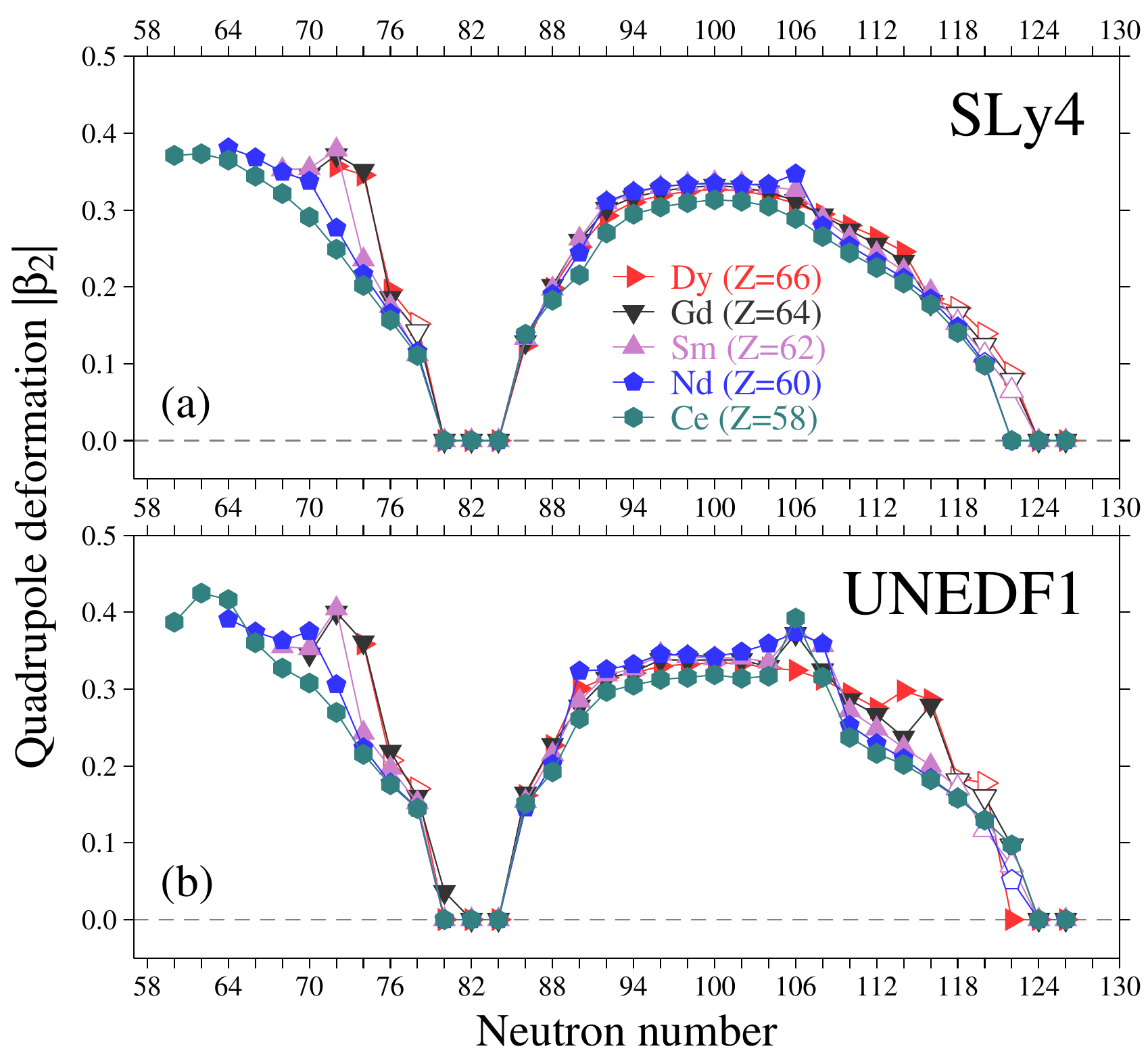}
\caption{Quadrupole deformation parameters $|\beta_2|$ of even-even nuclei from proton-dripline to $N=126$ in Ce, Nd, Sm, Gd, and Dy isotopes as a function of neutron number calculated with (a) SLy4 and (b) UNEDF1 interactions.
For visibility, oblate shapes ($\beta_2<0$) are represented by open symbols.
}
\label{SLy4.UNEDF1}
\end{center}
\end{figure}

\begin{figure}[!tb]
\begin{center}
\includegraphics[width=0.490\textwidth,keepaspectratio]{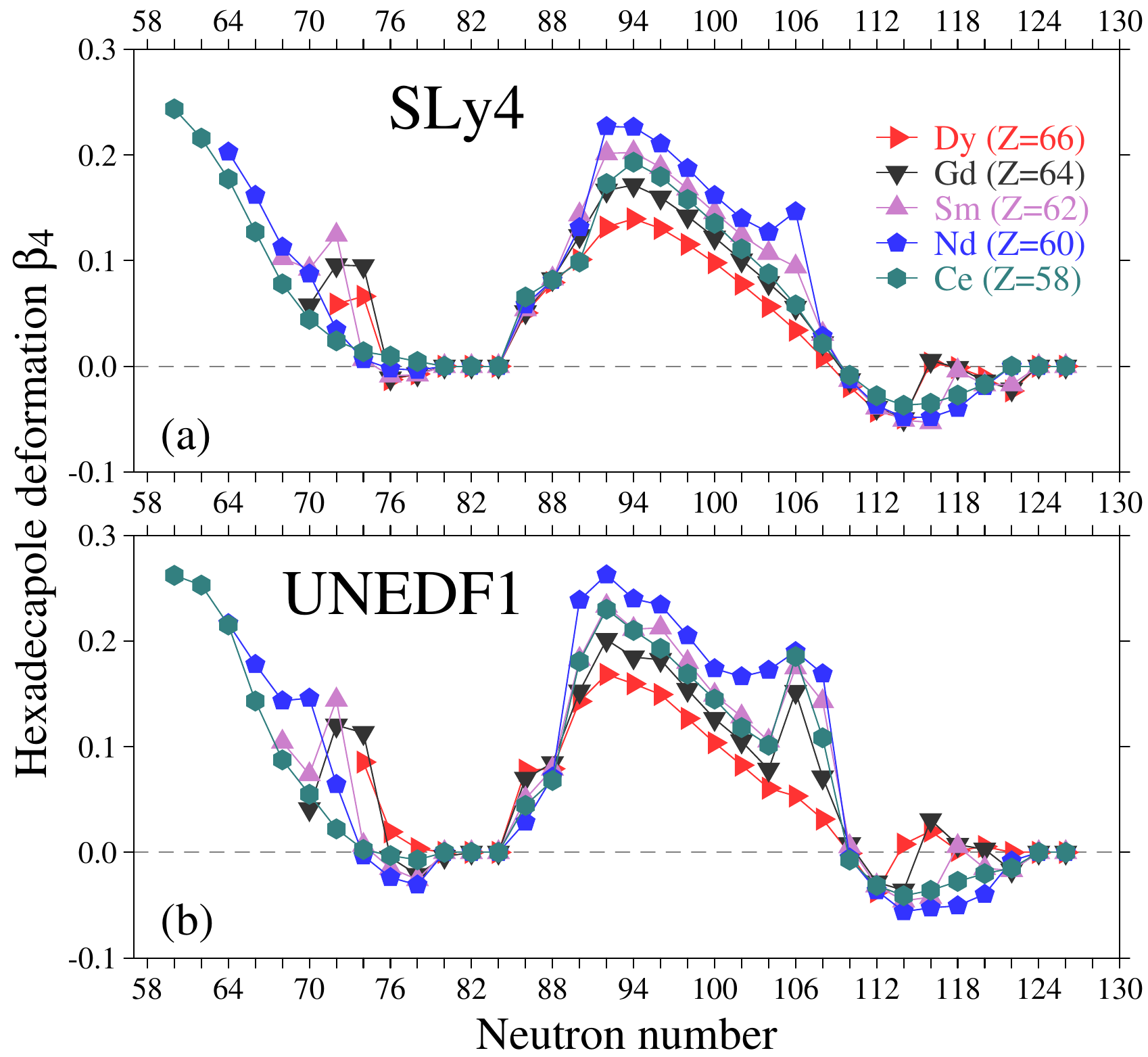}
\caption{Hexadecapole deformation parameters $\beta_4$ of even-even nuclei from proton-dripline to $N=126$ in Ce, Nd, Sm, Gd, and Dy isotopes as a function of neutron number calculated with (a) SLy4 and (b) UNEDF1 interactions.
}
\label{beta4-2}
\end{center}
\end{figure}

\begin{figure}[htb]
\begin{center}
\includegraphics[width=\columnwidth,keepaspectratio]{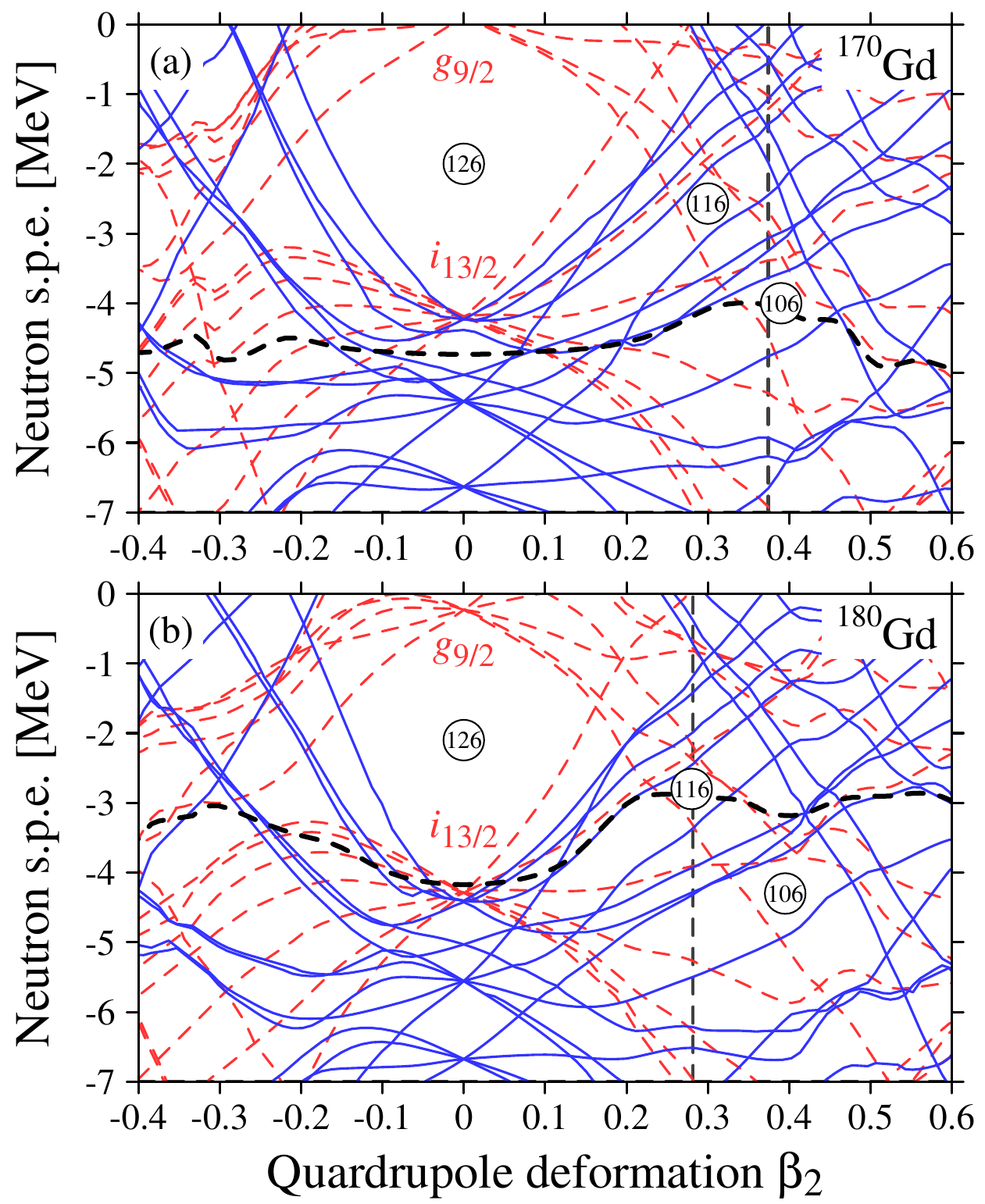}
\caption{Neutron single-particle energies (s.p.e) of (a) $^{170}$Gd and (b) $^{180}$Gd 
as a function of quadrupole deformation parameter $\beta_2$, calculated with UNEDF1 interaction.
}
\label{config.170Gd.180Gd}
\end{center}
\end{figure}

As we discussed in Sec.~III A, the
appearance of the ground state involving the second intruder orbit
strongly depends on the shell structure.
Here we investigate the interaction dependence of the present findings.
Figures~\ref{SLy4.UNEDF1} and ~\ref{beta4-2}, respectively, display the quadrupole and hexadecapole deformation parameters
of even-even nuclei of Ce ($Z=58$) to Dy ($Z=66$) isotopes, 
calculated with SLy4~\cite{Chabanat98}
and UNEDF1~\cite{Kortelainen12} interactions.
The SLy4 results predict the occupation of the
second intruder orbit in  $^{166}$Nd and $^{168}$Sm
at $N\approx 106$ where the sudden enhancement of the values of $\beta_2$ and $\beta_4$ is obtained, as in the SkM$^\ast$ case.
The UNEDF1 results predict the occupation in $^{164}$Ce, $^{166,168}$Nd,
$^{168,170}$Sm, and $^{170}$Gd.
For nuclei with $Z \approx 64$ and $N \approx 116$,
UNEDF1 produces highly-deformed ground states involving the occupation of
the second intruder [651]1/2 orbit from the spherical
$1g_{9/2}$ orbit, which is the same orbit to that in $^{166,168}$Nd,
as shown in Fig.~\ref{config.170Gd.180Gd}.
In the proton-rich region, both the SLy4 and UNEDF1 produce
the occupation of the second intruder orbit in $^{134}$Sm and $^{136,138}$Gd,
and proton-dripline nuclei of Dy isotopes.
Therefore, all the Skyrme interactions employed produce enlarged-deformed ground states 
in $^{138}$Gd, $^{166}$Nd, and $^{168}$Sm, indicating
the robust occupation of the second intruder orbit in this region.\\

\section{Summary}
\label{Summary}

We performed a systematic axial HFB (HFBTHO) survey with a common setup and focused on the region where the second intruder drives a highly-deformed state.\\

\textbf{Robust appearance of highly-deformed state.}
For representative nuclei in the second–intruder region, all three Skyrme interactions (SkM*, SLy4, UNEDF1) produce a common large prolate minimum at $\beta_2 \approx$ 0.35. 
An overlay figure shows that the curves almost overlap; the minimum appears in the same place with small quantitative differences.
This means the appearance of the highly-deformed state is robust against the interaction choice.

\textbf{Hexadecapole trends and coupling.}
At the prolate minimum, the extracted $\beta_4$ values are similar among the three interactions, and the curve shapes near the minimum are comparable. 
This points to a qualitatively robust $\beta_2$–$\beta_4$ coupling associated with the second–intruder configuration.

\textbf{Systematics and scope.}
Along rare–earth chains (e.g., Nd–Sm–Gd), we find the same qualitative picture: a common highly-deformed prolate minimum at $\beta_2 \approx$ 0.35 across SkM*, SLy4, and UNEDF1. 
A superheavy example, $^{272}$No, is shown separately as an illustrative case; we do not include it in the robustness because octupole and basis-size effects may change the ordering.

\textbf{Limitations.}
We used axial HFB with a fixed HO basis truncation and did not include octupole ($\beta_3$) or triaxial shapes. 
For very heavy nuclei the energy ordering of competing minima may change when $\beta_3$ and a larger basis are included.

\textbf{Outlook.}
The next step is a focused two–dimensional $(\beta_2,\beta_4)$ analysis for a few benchmark nuclei and the explicit inclusion of octupole ($\beta_3$) in heavy systems to fix the ordering when two minima compete within a few hundred keV. This will turn the present trend map into quantitative predictions.

\section*{Acknowledgments}

This work was in part supported by JSPS KAKENHI
Grants Nos. 23K22485, 25K07285, 25K01005,
and JSPS Bilateral Program No. JPJSBP120247715.

\end{document}